\documentclass[usenatbib,times]{mn2e}
\usepackage{graphicx}
\usepackage{epsf}

\def\xte{{\it RXTE}}

\def\H0{{\rm ~km~s^{-1}~Mpc^{-1}}}

\begin{document}

\title[Physical origin of the QPO and harmonic in BHBs]{An imperfect double: probing the physical origin of the low-frequency QPO and its harmonic in black hole binaries}

\author[Axelsson et al.]{Magnus Axelsson,$^{1,2}$\thanks{email: magnusa@astro.su.se}
Chris Done$^{3}$ and Linnea Hjalmarsdotter$^{4}$\\
$^{1}$Oskar Klein Center for CosmoParticle Physics, Department of Physics, Stockholm University, SE-106 91 Stockholm, Sweden\\
$^{2}$Department of Astronomy, Stockholm University, SE-106 91 Stockholm, Sweden\\
$^{3}$Department of Physics, Durham University, South Road, Durham DH1 3LE, UK\\
$^{4}$Sternberg Astronomical Institute, Moscow State University, Universitetskij pr. 13, 119899 Moscow, Russia\\
}

\date{Accepted --. Received --; in original form --}

\pagerange{\pageref{firstpage}--\pageref{lastpage}} \pubyear{2002}

\maketitle

\begin{abstract}

We extract the spectra of the strong low-frequency quasi-periodic
oscillation (QPO) and its harmonic during the rising phase of an
outburst in the black-hole binary XTE J1550-564. We compare these
frequency resolved spectra to the time-averaged spectrum and the
spectrum of the rapid ($<0.1$s) variability. The spectrum of the time
averaged emission can be described by a disc, a Compton upscattered
tail, and its reflection. The QPO spectrum is very similar to the
spectrum of the most rapid variability, implying it arises in the
innermost regions of the flow. It contains little detectable disc, and
its Compton spectrum is generally harder and shows less reflection
than in the time averaged emission. The harmonic likewise contains
little detectable disc component, but has a Compton spectrum which is
systematically softer than the QPO, softer even than the Compton tail
in the time averaged emission. We interpret these results in the
context of the truncated disc model, where the inner disc is replaced
by a hot flow. The QPO can arise in this picture from vertical
(Lense-Thirring) precession of the entire hot inner flow, and its
harmonic can be produced by the angular dependence of Compton
scattering within the hot flow. We extend these models to include
stratification of the hot flow, so that it is softer (lower optical
depth) at larger radii closer to the truncated disc, and harder
(higher optical depth) in the innermost parts of the flow where the
rapid variability is produced. The different optical depth with radius
gives rise to different angular dependence of the Comptonised
emission, weighting the fundamental to the inner parts of the hot
flow, and the harmonic to the outer. This is the first model which can
explain both the spectrum of the QPO, and its harmonic, in a self
consistent geometry.

\end{abstract}

\section{Introduction}

The correlated spectral and variability properties seen in accreting
black hole systems can be explained if the inner disc truncates at
some radius above the last stable orbit, evaporating into a hot inner
flow. Hot electrons in the inner flow up-scatter the soft disc
photons, giving rise to a Comptonised component in the
spectrum. Decreasing the transition radius as the mass accretion rate
increases from the low/hard to high/soft state gives both a
qualitative and quantitative description of the spectral softening
(both increase in disc component and increasing softness of the
Comptonised tail due to the larger cooling flux of seed photons from
the disc) and decrease in low frequency components of the power
spectral density \citep[e.g.,][]{dgk07}.

The lack of an inner disc means the flow can vertically
precess due to misalignment between the black hole spin and accretion
flow (Lense-Thirring precession). This is strongly differential for
single particle orbits, but for a viscously connected hot flow it can
result in global precession of the entire flow \citep{lub02}. This global 
precession is seen in numerical simulations
\citep{fra07}, and is a strong candidate to explain the origin of the
strong low-frequency quasi-periodic oscillation (QPO) seen in many
black hole binaries \citep{fra07,ing09}. \citet{ing09} show
that this scenario naturally explains the fact that the energy
spectrum of the QPO is dominated by the Comptonised emission
rather than the blackbody disk \citep[see, e.g.,][]{rm06}.

However, the structure of the hot flow is probably not simply a single
temperature, single optical depth Comptonising region. Cygnus X-1
shows clear evidence for two different regions in its Comptonised
spectrum close to the hard/soft spectral transition
\citep{yam13}. Spectral inhomogeneity is clearly required in order to
explain the complex pattern of time lags in the data \citep{mk89},
where slow variability stirred up at large radii (softer spectra) in
the hot flow propagates down to smaller radii, giving a lagged
modulation of a harder spectrum \citep{kot01,au06}. This correlation
of spectral shape with timescale is seen directly in Cyg X-1 using
frequency resolved spectra \citep{rev99}. This has an immediate
interpretation in the context of the truncated disc/hot inner flow
model as the outer parts of the hot flow are closer to the disc, so
intercept more seed photons so the spectrum should be softer than in
the inner regions of the flow which are more seed photon starved.

In \citet[][hereafter Paper I]{axe13} we used frequency resolved
spectroscopy to study the spectrum of the fastest variability through
a hard/soft transition. We found that it contained no contribution
from the disk blackbody, significantly less reflection and a harder
Comptonised spectrum than the time averaged continuum, in agreement
with the picture above of the innermost region of the hot flow. Here
we extend the frequency resolved spectral study to the QPO and its
harmonic in XTE~J1550-564.

Previous studies of the QPO spectrum have shown that both the rapid
variability and QPOs are tied predominantly to the Comptonised
emission rather than the disk \citep[e.g.,][]{chu01,gil03}.
However, \citet{sob06} did more detailed modelling of the QPO
spectrum, and were able to fit many of these using purely ionized
reflection rather than the direct Comptonised emission.  Here we use
the improved models of ionized reflection developed in Paper I to
reassess the contribution of reflection to the QPO spectrum,
especially as these models show that the very similar spectrum of the
fastest variability contains very little reflection (Paper I). We find
that the QPO spectra are better fit by Comptonisation models, with
little reflection, similar to the fastest variability spectra in Paper
I and consistent with being produced in the same innermost regions of
the hot flow.

The harmonic of the QPO is strong in the observations of XTE~J1550-564
considered here, with rms ranging from 0.3-0.5 times that of the
fundamental \citep{cui99}. Such a feature has also been seen in other
black hole transient systems, for example GX~339-4 \citep{bel05}.
There are no previous studies of the detailed spectrum of the
harmonic, though it is clear that it is systematically softer than the
QPO \citep{cui99}.  Recent work by \citet{vel13} has shown that the
existance of a strong harmonic is predicted by the angular radiation
pattern from Comptonisation of a precessing hot inner flow. We use
this model to interpret the harmonic spectra, and find that this is also
consistent with the inhomogeneous hot flow model discussed above, as the
harmonic should be stronger in the spectra from regions of lower
optical depth, weighting the harmonic to the outer (softer) regions of
the hot flow. This is the first model to explain both the QPO and its
harmonic in the same inhomogeneous hot flow geometry as required to
explain the time lags.

\section{Data analysis}
\label{analysis}

In this study we use archival data of XTE~J1550-564 from the
Proportional Counter Array \citep[PCA;][]{jah96} instrument onboard
the {\xte} satellite. The observations are made in the time period
1998-09-09 to 1998-09-16 (MJD 51065 to 51072) and cover the rise of
the strong initial flare of the 1998 outburst \citep[for more details
on this outburst see][]{smi98}.

To extract the spectra of the QPO and harmonic, we follow the same
procedure as in Paper I. In every observation we extracted a light
curve for each available channel and constructed a power density
spectrum. To determine the contribution of the QPO/harmonic we fit
these features using Lorentzian functions, which were then integrated
to produce the rms variability. Finally, combining the results for all
channels allows us to build the energy spectrum of the QPO/harmonic.

As in Paper I, we follow the standard approach and add a systematic
error of 1 per cent to all bins. We note that this leads to the
systematics dominating the uncertainties for the average spectra, as
seen by the extremely low values of $\chi^2/$dof. 
In Paper I, the additional errors introduced via
the power spectrum meant that systematics did not dominate the rms
spectra of the fastest ($>10$~Hz) variability. However, the QPO 
has more variability power, so the QPO spectra are better determined
so systematics can dominate the QPO spectra.

The data were fit in \textsc{xspec} using the same spectral model as
in Paper I: a disc blackbody \citep[{\sc diskbb};][]{mit84}, thermal
Comptonization \citep[{\sc nthcomp};][]{zyc99} with seed photons tied
to the inner disc temperature, and reflection of the Compton spectrum
({\sc rfxconv}, based on the \citealp{rf05} ionised reflection tables,
but re-coded into a convolution form so they can be used on any
continuum; \citealp{koh11}). We note that this reflection model is
more physical than the {\sc pexriv} based ionised reflection models
used by \cite{sob06} as it includes Comptonisation within the disc
photosphere itself which broadens the line and edge features
\citep{you01}. We fix the inclination at $70^\circ$, as determined
from the orbital elements of the system \citep{oro02}. We also include 
an absorption line ({\sc gabs}), in order to model absorption through 
a wind from the disc as is generally seen in high inclination systems
\citep{pon12}.

\section{Results}
\label{results}

Table~\ref{datatable} shows the results of fitting our spectral model
to the data. For comparison, we also include the fits of the rapid
variability and continuum spectrum from Paper I. The model above gives
an adequate fit $\chi^2$/dof$\sim 1$ to both the QPO and harmonic data.

\begin{table*}
\tabcolsep=0.12cm
\begin{tabular}{l l l l l l l l l l l}
ObsID & $T_{\rm bb}$ & $N_{\rm bb}$ & $\Gamma$ & $kT_{\rm e}$ &  $R$ & $E_{\rm line}$ & EW & $\log\xi$ & $\chi^2/dof$& Flux \\
& {\scriptsize (keV)} & & & {\scriptsize (keV)} & & {\scriptsize (keV)} & {(\scriptsize $\times 10^{-2}$ keV)} & & & {\scriptsize erg cm$^{-2}$ s$^{-1}$}\\
\hline

30188-06-01-00  & $0.56^{+0.31}_{-0.26}$ & $200^{+570}_{-200}$ & $1.77^{+0.03}_{-0.10}$ & $14.8^{+7.3}_{-3.0}$ & $0.44^{+0.12}_{-0.11}$ & $6.84^{+0.11}_{-0.11}$ & $9.5^{+4.0}_{-2.7}$ & $2.80^{+0.03}_{-0.09}$  & 2.3/36 & $4.4\times10^{-8}$ \\ [1ex]
\hspace{2mm} rms  & & & $1.69^{+0.03}_{-0.03}$ & & $0.07^{+0.09}_{-0.07}$ & & & & 17.4/54 & \\ [1ex]
\hspace{2mm} qpo & & $0^{+360}_{-0}$ & $1.72^{+0.02}_{-0.02}$ & & $0.06^{+0.12}_{-0.06}$ & & & & 5.9/54& \\ [1ex]
\hspace{2mm} harmonic & & $0^{+200}_{-0}$ & $1.92^{+0.05}_{-0.06}$ & & $0.50^{+0.28}_{-0.23}$ & & & & 5.8/49& \\ [1ex]

30188-06-01-01 & $0.55^{+0.04}_{-0.04}$ & $260^{+590}_{-260}$ & $1.80^{+0.03}_{-0.04}$ & $15.1^{+5.8}_{-3.0}$ & $0.43^{+0.14}_{-0.11}$ & $6.85^{+0.09}_{-0.09}$ & $9.2^{+3.5}_{-3.6}$ & $2.80^{+0.67}_{-0.09}$ & 7.0/36 & $4.0\times10^{-8}$ \\ [1ex]
\hspace{2mm} rms & & &$1.66^{+0.09}_{-0.08}$ & & $0.06^{+0.25}_{-0.06}$ & & & & 50.8/54 & \\ [1ex]
\hspace{2mm} qpo & & $220^{+660}_{-220}$ & $1.80^{+0.03}_{-0.03}$ & & $0.23^{+0.18}_{-0.16}$ & & & & 23.4/54& \\ [1ex]
\hspace{2mm} harmonic & & $0^{+230}_{-0}$ & $1.95^{+0.03}_{-0.05}$ & & $0.01^{+0.10}_{-0.01}$ & & & & 24.0/49& \\ [1ex]

30188-06-01-02 & $0.55^{+0.28}_{-0.08}$ & $410^{+1100}_{-410}$ & $1.89^{+0.02}_{-0.02}$ & $10.6^{+2.4}_{-1.5}$ & $0.56^{+0.15}_{-0.14}$ & $6.83^{+0.10}_{-0.09}$ & $11^{+5.0}_{-4.7}$ & $2.78^{+0.60}_{-0.09}$ & 4.5/36 & $4.3\times10^{-8}$  \\ [1ex]
\hspace{2mm} rms & & & $1.78^{+0.03}_{-0.03}$ & & $0.0^{+0.07}_{-0.0}$ & & & & 35.2/54& \\ [1ex]
\hspace{2mm} qpo & & $0^{+63}_{-0}$ & $1.79^{+0.01}_{-0.01}$ & & $0.22^{+0.07}_{-0.07}$ & & & & 55.4/54& \\ [1ex]
\hspace{2mm} harmonic & & $0^{+67}_{-0}$ & $2.07^{+0.03}_{-0.03}$ & & $0.27^{+0.12}_{-0.11}$ & & & & 43.9/47& \\ [1ex]

30188-06-01-03 & $0.53^{+0.25}_{-0.09}$ & $970^{+1100}_{-970}$ & $1.94^{+0.03}_{-0.09}$ & $10.9^{+2.3}_{-1.6}$ & $0.52^{+0.15}_{-0.16}$ & $6.80^{+0.08}_{-0.08}$ & $9.6^{+2.5}_{-3.7}$ & $2.78^{+0.70}_{-0.08}$ & 7.0/36 & $4.4\times10^{-8}$ \\ [1ex]
\hspace{2mm} rms & & & $1.79^{+0.03}_{-0.05}$ & &  $0.07^{+0.21}_{-0.07}$ & & & & 17.8/54 & \\ [1ex]
\hspace{2mm} qpo & & $0^{+250}_{-0}$ & $1.76^{+0.03}_{-0.03}$ & & $0.33^{+0.17}_{-0.14}$ & & & & 41.2/54& \\ [1ex]
\hspace{2mm} harmonic & & $0^{+182}_{-0}$ & $2.03^{+0.04}_{-0.04}$ & & $0.13^{+0.14}_{-0.13}$ & & & & 23.1/47& \\ [1ex]

30188-06-04-00 & $0.53^{+0.19}_{-0.53}$ & $4100^{+1200}_{-2000}$ & $2.03^{+0.03}_{-0.03}$ & $9.9^{+1.7}_{-0.7}$ & $0.54^{+0.17}_{-0.15}$ & $6.82^{+0.09}_{-0.09}$ & $9.6^{+2.9}_{-3.9}$ & $2.79^{+0.68}_{-0.05}$ & 6.1/36 & $5.1\times10^{-8}$  \\ [1ex]
\hspace{2mm} rms & & & $1.90^{+0.03}_{-0.03}$ & & $0.33^{+0.12}_{-0.11}$ & & & & 38.9/54 & \\ [1ex]
\hspace{2mm} qpo & & $0^{+90}_{-0}$ & $1.87^{+0.01}_{-0.01}$ & & $0.43^{+0.10}_{-0.10}$ & & & & 48.7/54& \\ [1ex]
\hspace{2mm} harmonic & & $0^{+86}_{-0}$ & $2.25^{+0.03}_{-0.03}$ & & $0.42^{+0.13}_{-0.12}$ & & & & 40.7/47& \\ [1ex]

30188-06-05-00 & $0.58^{+0.08}_{-0.58}$ & $8400^{+880}_{-2800}$ & $2.19^{+0.07}_{-0.04}$ & $11.6^{+3.3}_{-1.4}$ & $0.54^{+0.20}_{-0.17}$ & $6.81^{+0.10}_{-0.11}$ & $9.0^{+3.9}_{-2.9}$ & $2.81^{+0.18}_{-0.22}$& 6.0/36 & $6.0\times10^{-8}$ \\ [1ex]
\hspace{2mm} rms & & & $2.01^{+0.03}_{-0.03}$ & & $0.10^{+0.10}_{-0.10}$ & & & &  28.2/54 & \\ [1ex]
\hspace{2mm} qpo & & $0^{+51}_{-0}$ & $2.03^{+0.01}_{-0.01}$ & & $0.48^{+0.12}_{-0.11}$ & & & & 84.2/54& \\ [1ex]
\hspace{2mm} harmonic & & $400^{+450}_{-370}$ & $2.33^{+0.06}_{-0.06}$ & & $0.10^{+0.23}_{-0.10}$ & & & & 39.9/47& \\ [1ex]

30188-06-06-00 & $0.65^{+0.08}_{-0.45}$ & $9100^{+540}_{-1000}$ & $2.33^{+0.06}_{-0.06}$ & $11.5^{+4.5}_{-1.5}$ & $0.70^{+0.19}_{-0.20}$ & $6.80^{+0.08}_{-0.09}$ & $11^{+2.7}_{-3.7}$ & $2.75^{+0.17}_{-0.16}$ & 5.6/36 & $5.4\times10^{-8}$\\ [1ex]
\hspace{2mm} rms  & & & $2.07^{+0.02}_{-0.04}$ & & $0.30^{+0.25}_{-0.13}$ & & & & 38.7/54 & \\ [1ex]
\hspace{2mm} qpo & & $0^{+22}_{-0}$ & $2.14^{+0.01}_{-0.01}$ & & $0.54^{+0.06}_{-0.10}$ & & & & 111/54& \\ [1ex]
\hspace{2mm} harmonic & & $340^{+230}_{-190}$ & $2.51^{+0.08}_{-0.08}$ & & $0.59^{+0.21}_{-0.19}$ & & & & 6.4/47& \\ [1ex]

30188-06-07-00 & $0.70^{+0.07}_{-0.51}$ & $6600^{+350}_{-560}$ & $2.27^{+0.08}_{-0.06}$ & $10.4^{+2.7}_{-1.0}$ & $0.64^{+0.15}_{-0.20}$ & $6.81^{+0.08}_{-0.08}$ & $11^{+3.0}_{-3.4}$ & $2.78^{+0.23}_{-0.22}$ & 6.3/36 & $6.4\times10^{-8}$ \\ [1ex]
\hspace{2mm} rms & & & $2.11^{+0.03}_{-0.03}$ & & $0.29^{+0.08}_{-0.11}$ & & & & 35.8/54 & \\ [1ex]
\hspace{2mm} qpo & & $0^{+38}_{-0}$ & $2.12^{+0.01}_{-0.01}$ & & $0.46^{+0.10}_{-0.09}$ & & & & 36/54& \\ [1ex]
\hspace{2mm} harmonic & & $0^{+24}_{-0}$ & $2.35^{+0.03}_{-0.03}$ & & $0.49^{+0.10}_{-0.10}$ & & & & 16/47& \\ [1ex]

30188-06-09-00 & $0.66^{+0.08}_{-0.45}$ & $9600^{+480}_{-900}$ & $2.39^{+0.08}_{-0.05}$ & $12.8^{+5.3}_{-1.9}$ & $0.67^{+0.16}_{-0.20}$ & $6.79^{+0.08}_{-0.08}$ & $9.5^{+3.0}_{-3.1}$ & $2.73^{+0.14}_{-0.16}$ & 6.4/36 & $6.7\times10^{-8}$  \\ [1ex]
\hspace{2mm} rms & & & $2.08^{+0.03}_{-0.03}$ & & $0.22^{+0.13}_{-0.12}$ & & & & 21.8/54 & \\ [1ex]
\hspace{2mm} qpo & & $0^{+78}_{-0}$ & $2.24^{+0.01}_{-0.01}$ & & $0.31^{+0.09}_{-0.09}$ & & & & 69.5/54 & \\ [1ex]
\hspace{2mm} harmonic & & $0^{+37}_{-0}$ & $2.48^{+0.03}_{-0.03}$ & & $0.50^{+0.13}_{-0.12}$ & & & & 25.8/47& \\ [1ex]

\end{tabular}
\caption{Fit results for the model \textsc{tbabs$\times$gabs$\times$(diskbb+nthcomp+nthcomp$\times$rfxconv)}. Errors indicate 90\% confidence intervals.}
\label{datatable}
\end{table*}

\subsection{Spectrum of QPO and rapid variability}

\begin{figure*}
\includegraphics[width=17cm,angle=0]{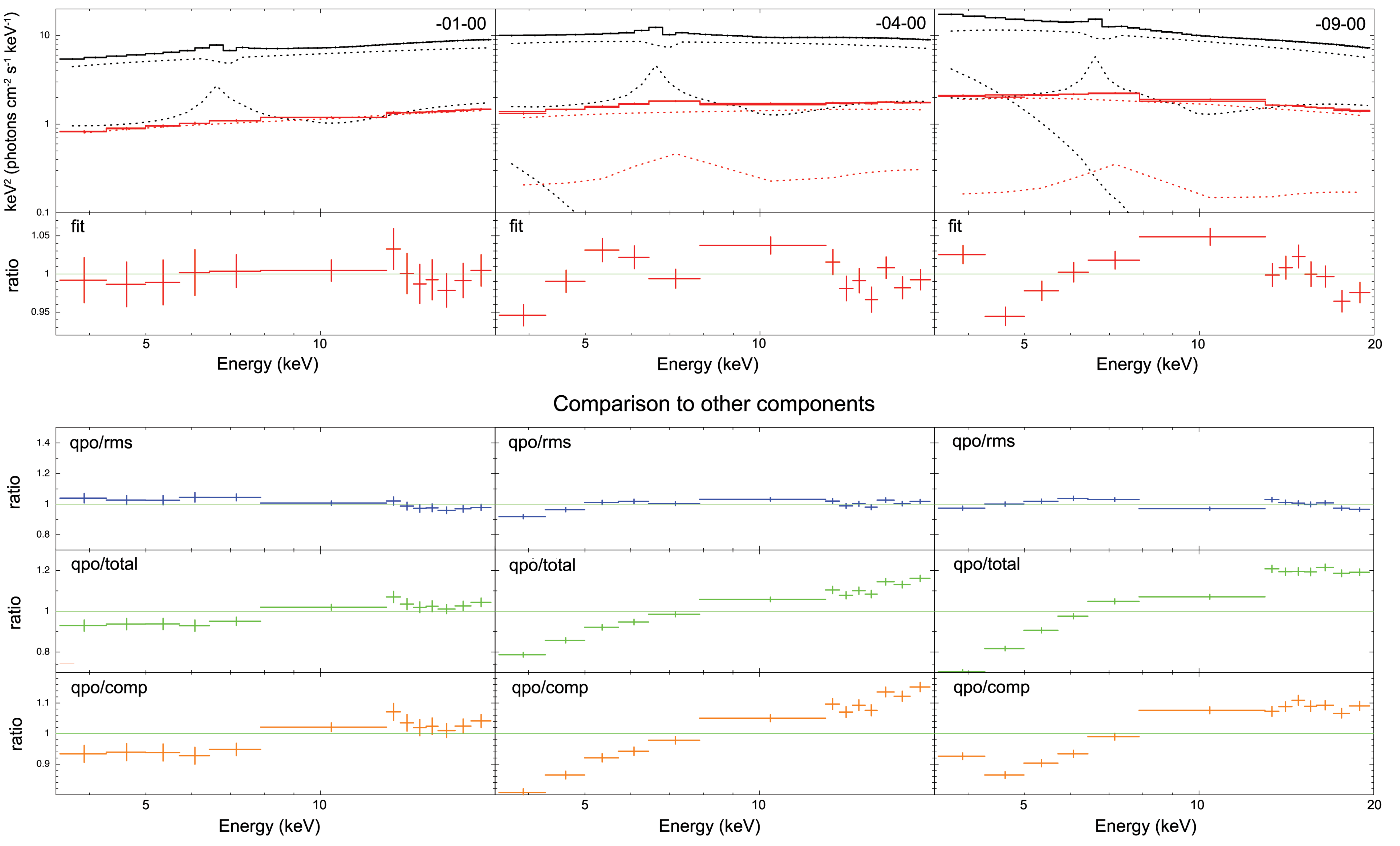}
\caption{Top panel shows fits to the time averaged spectrum (black)
and QPO (red). The second panel shows the ratio of the QPO spectrum (red) to
the model fit in the upper panel, showing that the model describes the
data well. Further down, the third panel shows the ratio between the QPO spectrum
and the best-fit model for the rapid variability (blue), showing
their similarity. The fourth panel shows the ratio of the QPO spectrum
(green) to the time averaged continuum. There is a clear drop at low
energies when the disc dominates, showing that the QPO does not
modulate the disc emission. The bottom panel shows the ratio when
comparing the QPO spectrum to only the Comptonised component from the
time averaged continuum (orange). The QPO is subtly harder than the
time averaged Comptonised emission, by an increasing amount as the
spectrum softens.}
\label{QPOfits}
\end{figure*}

The full details of the fits to all spectra through the transition are
given in Table 1, while Fig.~\ref{QPOfits} shows the time averaged
(total) continuum and spectrum of the QPO at the start, midpoint and
end of the hard/soft transition. The time averaged Compton spectrum
(black points in the top panel) softens dramatically as the source
makes the transition, with an increasing amount of disc and disc
reflection present. The QPO spectrum (red points in the top panel) is
fit with the same model components, but with the spectral index of the
Comptonised emission, and the amount of disc and reflection allowed to
vary. The next lowest panel for each ObsID shows the QPO spectral
model residuals (red points). 

Below the fit results, we also present three panels where we compare
the best-fit results of the QPO  to other components. The first panel shows
the ratio of  QPO spectrum to the model fit to the rapid variability 
($>10$~Hz: blue points from Paper I). The QPO spectrum is always similar 
to that of the rapid variability throughout the outburst. Thus we expect (and
see) similar parameters in the fits in Table 1 as for the spectrum of
the rapid variability in Paper I. This shows that the QPO spectrum is well
described by thermal Comptonisation with only a small amount of
reflection, in contrast to the reflection dominated fits of
\cite{sob06}, due to our use of more sophisticated reflection models
(see Paper I). The lack of a strong disc component in the QPO spectrum can
clearly be seen in the next panel down, which shows the ratio of the
QPO spectrum to the model fit to the time averaged data (total
spectrum: green points). There is a clear drop in this ratio at low
energies where the disc component contributes.

The fact that there is less disc and less reflection in the QPO
spectrum means that to zeroth order it is well described by the
thermal Compton emission alone. However, more subtle effects can be
seen in the bottom panel, which shows the ratio of the QPO spectrum to
the model of thermal Comptonisation fit to the time averaged spectrum
(orange points). The QPO spectrum evolves from being consistent with
the Comptonised emission of the time averaged spectrum in the hardest
states, to being increasingly harder than the time averaged spectrum
as the source makes the transition, which is indeed the same behaviour
as was seen in the rapid variability spectrum of Paper I, making it
very likely that the QPO and rapid variability arise in the same
region of the accretion flow.

\subsection{The spectrum of the harmonic}

\begin{figure*}
\includegraphics[width=17cm,angle=0]{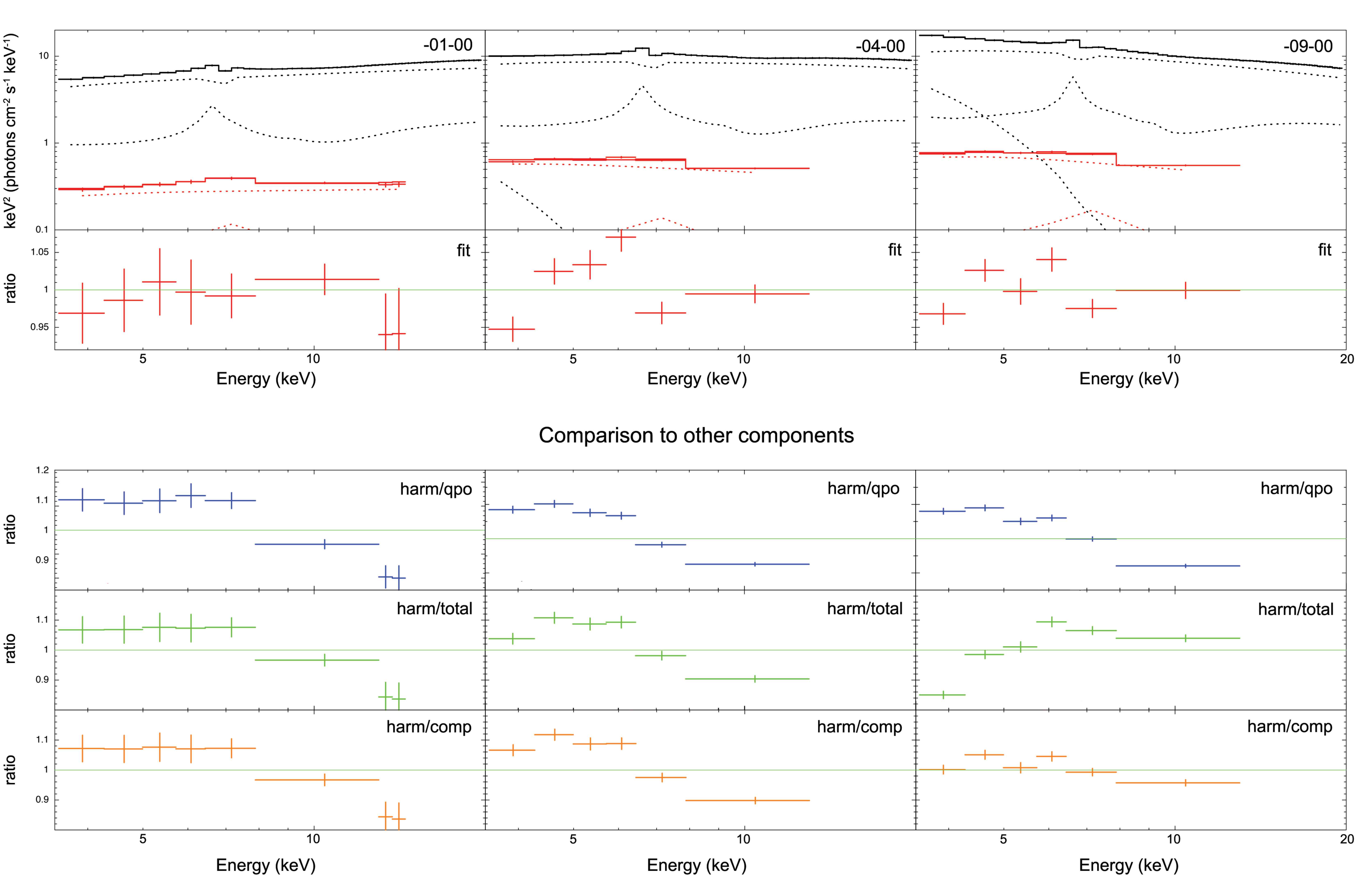}
\caption{Fit of continuum spectrum and harmonic (top panel for each fit) along with residuals of the fit (red). Lower panels in each spectrum show 
the ratio between the harmonic data and best-fit models for the QPO (blue) and continuum (green). The harmonic is noticeably 
softer than the QPO and is not so different from the total spectrum, although there is still a drop at energies corresponding to the disc
emission, signalling that the direct disc emission is not being modulated. The bottom panel shows the ratio when comparing the spectrum of the harmonic 
to only the Comptonised component from the continuum (orange).}
\label{harmfits}
\end{figure*}

Figure~\ref{harmfits} shows the same spectra as in Fig.~\ref{QPOfits},
with the vertical panels showing the same sequence of comparisons
except for this figure we select the harmonic of the QPO rather than
the fundamental. In agreement with the results found by \cite{cui99},
we find that the harmonic becomes very weak at higher energies, and we
are unable to constrain it in the power spectrum above $\sim12$
Hz. Interestingly, in the hardest observations it is seen up to
somewhat higher energies (first two spectra in Fig.~\ref{harmfits}).

The harmonic is well fit by the same model components (red points, top
and second panel), but with very different parameters (Table 1). The
harmonic is noticeably steeper than the rapid variability (blue
points), and is not so different to the total spectrum (green points),
although there is still a drop at energies corresponding to the disc
emission, again signalling that the direct disc emission is not being
modulated (see also Table 1).  Instead, the ratio of the harmonic to
the Compton spectrum alone (orange, bottom panel) shows that the
harmonic is systematically softer than the Comptonised component of
the time averaged spectrum. More subtly, Table 1 also hints that
there could be more reflection in the harmonic than in the QPO. 

Thus it is clear that the harmonic arises in a different environment
than the QPO. The fact that the harmonic spectrum is softer than the
QPO and shows stronger signs of reflection points to the harmonic
arising further out in the flow, in a region seeing more soft seed
photons and where the disc subtends a larger solid angle.

\subsection{Limitations of the spectral fits}

While the fit statistics show that the fits are generally good, the
residuals of the fits in the softer states hint that there may still
be details not fully captured by our model. We tied the Comptonisation
seed photon temperature to be the same as that of the observed disc
component seen in the average spectrum, yet the suggested geometry for
the softer states has the inner disc extending underneath the outer
parts of the corona.  The corona is translucent, with optical depth
$\sim 1-2$ so photons from the disc underneath the corona are not
observed as they are all Compton scattered. Thus these form the
predominant source of seed photons, and are higher temperature than
the observed disc emission which comes from further out in the disc
(see Done \& Kubota 2006 for a detailed discussion of this). We can
find better fits by allowing the seed photon temperature in the corona
to be higher than that of the observed disc emission. However, this
seed photon temperature cannot be easily constrained from the time
average spectrum alone as the observed disc overlaps the energy
band of the downturn. The lack of a strong disc component in the QPO 
spectra means that these are more sensitive to the seed photon energy 
of the Comptonised emission than the corresponding time average
spectra. However, the QPO spectra are of limited quality - there are
only 13 data points in each spectrum - and hence they cannot reasonably
constrain the seed photon temperature as well as the amount of
reflection, Compton spectral index and normalisation of the disc and
Compton components.

We explore possible consequences of this limitation by stepwise increasing
the seed photon temperature for the time averaged and QPO spectra (always 
using the same value for both datasets). We find that this indeed allows 
us to find fits where the residuals seen in Fig.~\ref{QPOfits} mostly disappear. 
The resulting parameters show very similar trends to those presented 
above, in that the Compton spectrum of the QPO is generally harder, 
and has less reflection than that of the time average spectrum.

\section{Discussion}
\label{discussion}

Although there have been many proposed mechanisms for QPOs in
black-hole binaries, these have focussed mainly on how to match the
observed characteristic frequencies rather than the spectrum of the
QPO. \citet{ing09} showed that vertical precession of the entire hot
inner flow, as seen in the numerical simulations of \citet{fra07},
could provide a mechanism for both the frequency and spectrum of the
QPO. Precession of the hot flow imprints the modulation onto the
spectrum of the hot flow, naturally explaining the similarity between
the QPO spectrum and the Comptonised emission in the continuum
spectrum, and the lack of disc in the QPO. 

This model can also be extended to encompass the broadband variability
as well as the QPO \citep{ing11,ing12}, by propagating
fluctuations down through the entire precessing hot flow. This gives a
framework in which to interpret the more subtle features seen in the
QPO spectrum, and in the spectra of the most rapid variability (Paper
I). The most rapid variability is only produced in the very inner
region of the flow, while slower variability is produced at larger
radii, and propagates down to the inner regions.  All the hot flow
within the truncated disc can precess vertically, but the regions
furthest from the disc (i.e., closer to the centre) have harder spectra
and produce less reflection as they subtend a smaller solid angle to
the disc and are shielded from the disc by the radial optical depth of
the rest of the flow. In the softest spectra seen here, the disc
probably extends underneath the outer parts of the corona, preventing
these regions from vertically precessing. Only the very innermost
parts of the flow can precess, and these have the fastest stochastic
variability and the hardest spectra. 

This picture then explains the similarlity of the QPO and rapid
variability. However, it can also explain the difference in spectrum
of the QPO and its harmonic. The harmonic is clearly softer than the
QPO, but again this is mostly a change in the Compton continuum shape
rather than in the amount of disc emission as the harmonic 
contains little detectable disc component.

Instead, in the context of a Compton scattering slab viewed at an angle
$\theta$, the observed flux can be well approximated by \citep{pss83,vp04}
\begin{equation}
F_{E}(\theta)\propto I(\theta) \cos{\theta}\approx
(1+b\cos\theta)\cos\theta
=b/2+\cos\theta+b/2\cos2\theta \, .
\end{equation}
This approximation is fairly good even when $I(\theta)$ is not well
described by $(1+b\cos\theta)$ \citep[J. Poutanen, priv. comm., see
also][]{pss83,pg03}.  This means that the relative strength of the
harmonic to QPO is set by $|b|$, which is dependent on the optical
depth \citep{pss83,pg03}. Lower optical depths give rise to larger
$|b|$, so produce stronger harmonics \citep[see, e.g., fig. 2
in][]{vp04}.

The optical depth and/or temperature must increase with radius in the
inhomogeneous picture for the Comptonised region developed to explain
the spectral lags, so that harder spectra come from the innermost
regions \citep{kot01,au06}. Our spectral fits
assumed that the electron temperature remains constant throughout the
region (i.e., that only the optical depth changes to make the change in
spectral index). Thus in our fits the optical depth for the outer
regions must be lower than in the harder, innermost regions, and thus
the softer outer regions of the flow will contribute more to the
harmonic spectrum. This means that the harmonic has a stronger
contribution from the parts of the flow closest to the disc, matching
the observational results of a softer spectrum and higher reflection.

A more quantitative result would require constraining the electron
temperature in the inner and outer regions independently of the
optical depth, which is beyond the current data. It should also depend
on the scale height and/or detailed shape of the precessing flow.
Equation 1 is strictly valid only for a slab, whereas we expect that
the hot flow has an appreciable scale height and shape which also has
a role in modulating the emission \citep{ing12}. It also assumes that
relativistic effects are negligible (i.e., typical radii are some
$\sim$ tens of Schwarzschild radii). Closer to the black hole, effects
such as stronger gravitational curvature and faster Keplerian motion
cause the dependence of the flux on $\cos{\theta}$ to be weaker
\citep{pg03}. Together these effects could make substantial
differences to the very simple model above, but the principle of a
stronger contribution to the harmonic in the outer regions should
remain.

Finally, we note that the Comptonised emission is expected to be polarized
\citep{pss83,vp04}, and a precessing flow would mean that a QPO will be present 
in the polarization signal as well. Such a detection would therefore give extremely 
strong support to the Lense-Thirring origin of the QPO.

\section{Summary and Conclusions}

We have presented the first systematic analysis of the QPO spectrum
through a hard/soft transition. We confirm previous results that it
contains little detectable disc component, and for the first time show
that its Compton spectrum is generally harder than in the time
averaged emission, with less reflection. This makes it very similar to
the spectrum of the rapid ($>10$ Hz) variability. Additionally, for
the first time, we show the detailed spectrum of the harmonic.  Like
the QPO, this shows little detectable disc component, but unlike the
QPO it shows Comptonised emission which is softer than the
Comptonisation component in the time-averaged continuum.  The results
are matched by expectations from the recently proposed scenario where
the QPO is caused by Lense-Thirring precession of the entire hot inner
flow. This model {\em predicts} the existance of harmonic structure to
the QPO due to the angular distribution of the Comptonised emission
coupled to the precession. The difference in spectra between the QPO
and its harmonic can be additionally explained if the flow is radially
stratified, with optical depth increasing towards the inner regions of
the flow as this weights the harmonic towards the outer regions of the
flow, where the spectrum is softer and produces more reflection.  This
framework is thus the first physical model encompassing both the QPO
and harmonic.

\section*{Acknowledgments}
This work was supported by The Royal Swedish Academy of Sciences and has made use of data obtained through the High Energy 
Astrophysics Science Archive Research Center (HEASARC) Online Service, provided by NASA/Goddard Space Flight Center. We 
thank Adam Ingram and Juri Poutanen for helpful discussions.


\begin{thebibliography}{}
 
\bibitem[Ar{\'e}valo \& Uttley(2006)]{au06} Ar{\'e}valo P., \& Uttley P.,\ 2006, MNRAS, 367, 801
\bibitem[Axelsson et al.(2013)]{axe13} Axelsson M., Hjalmarsdotter L. Done, C., 2013, MNRAS, 431, 1987
\bibitem[Belloni et al.(2005)]{bel05} Belloni T., Homan J., Casella P. et al., 2005, A\&A, 440, 207 
\bibitem[Churazov et al.(2001)]{chu01} Churazov E., Gilfanov M., Revnivtsev M.,\ 2001, MNRAS, 321, 759 
\bibitem[Cui et al.(1999)]{cui99} Cui W., Zhang S., Chen W., Morgan. E., 1999, ApJ, 512, L43 
\bibitem[Done \& Kubota(2006)]{dk06} Done C., \& Kubota A.,\ 2006, MNRAS, 371, 1216 
\bibitem[\protect\citeauthoryear{Done, Gierli{\'n}ski \& Kubota}{2007}]{dgk07} Done C., Gierli{\'n}ski M., Kubota A., 2007, A\&ARv, 15, 1 
\bibitem[Fragile et al.(2007)]{fra07} Fragile C.~P., Blaes O.~M., Anninos P., Salomonson J.~D., 2007, ApJ, 668, 417
\bibitem[Gilfanov et al.(2003)]{gil03} Gilfanov M., Revnivtsev M., Molkov S.,\ 2003, A\&A, 410, 217 
\bibitem[Ingram et al.(2009)]{ing09} Ingram A., Done C., Fragile P.~C.,\ 2009, MNRAS, 397, L101 
\bibitem[Ingram \& Done(2011)]{ing11} Ingram A. \& Done C., 2011, MNRAS, 415, 2323
\bibitem[Ingram \& Done(2012)]{ing12} Ingram A. \& Done C.,\ 2012, MNRAS, 419, 2369 
\bibitem[Jahoda et al.(1996)]{jah96} Jahoda K., Swank J.~H., Giles A.~B. et al., 996, Proc. SPIE, 2808, 59
\bibitem[Kohlemainen, Done \& D{\`i}az Trigo(2011)]{koh11} Kohlemainen M., Done C., D{\`i}az Trigo M., 2011, MNRAS, 416, 311
\bibitem[Kotov et al.(2001)]{kot01} Kotov O., Churazov E., Gilfanov M.,\ 2001, MNRAS, 327, 799
\bibitem[Lubow et al.(2002)]{lub02} Lubow S.~H., Ogilvie G.~I., \& Pringle J.~E.,\ 2002, MNRAS, 337, 706 
\bibitem[Mitsuda et al.(1984)]{mit84} Mitsuda K., Inoue H., Koyama K. et al.,\ 1984, PASJ, 36, 741 
\bibitem[Miyamoto \& Kitamoto(1989)]{mk89} Miyamoto S., \& Kitamoto S.,\ 1989, Nature, 342, 773
\bibitem[Orosz et al.(2002)]{oro02} Orosz J.~A., Groot P.~J., van der Klis M. et al.,\ 2002, ApJ, 568, 845 
\bibitem[Ponti et al.(2012)]{pon12} Ponti G., Fender R.~P., Begelman M.~C. et al.\ 2012, MNRAS, 422, L11
\bibitem[Pozdnyakov, Sobol \& Sunyaev(1983)]{pss83} Pozdnyakov L.~A., Sobol I.~M., Syunyaev R.~A.,\ 1983, Astrophysics and Space Physics Reviews, 2, 189 
\bibitem[Poutanen \& Gierli{\'n}ski(2003)]{pg03} Poutanen J., \& Gierli{\'n}ski M.\ 2003, MNRAS, 343, 1301 
\bibitem[Remillard \& McClintock(2006)]{rm06} Remillard R.~A., \& McClintock J.~E.,\ 2006, ARA\&A, 44, 49
\bibitem[\protect\citeauthoryear{Revnivtsev et al.}{1999}]{rev99} Revnivstev M., Gilfanov M., Churazov E., 1999, A\&A, 347, L23
\bibitem[Ross \& Fabian(2005)]{rf05} Ross R.~R., Fabian A.~C.,\ 2005, MNRAS, 358, 211 
\bibitem[Smith(1998)]{smi98} Smith D. A., 1998, IAU Circ. 7008
\bibitem[Sobolewska \& {\.Z}ycki(2006)]{sob06} Sobolewska M. A., {\.Z}ycki P. T., 2006, MNRAS, 370,405
\bibitem[Veledina et al.(2013a)]{vel12} Veledina A., Poutanen J., Vurm I.,\ 2013, MNRAS, 430, 3196 
\bibitem[Veledina et al.(2013b)]{vel13} Veledina A., Poutanen J., Ingram A., 2013, in prep.
\bibitem[Viironen \& Poutanen(2004)]{vp04} Viironen K., \& Poutanen J.,\ 2004, A\&A, 426, 985 
\bibitem[Yamada et al.(2013)]{yam13} Yamada S., Makishima K., Done C. et al.,\ 2013, arXiv:1304.1968 
\bibitem[Young et al.(2001)]{you01} Young A.~J., Fabian A.~C., Ross R.~R., Tanaka Y.,\ 2001, MNRAS, 325, 1045 
\bibitem[{\.Z}ycki, Done \& Smith(1999)]{zyc99} {\.Z}ycki P., Done C., Smith D.A., 1999, MNRAS, 309, 561

\end{thebibliography}
\end{document}